\documentclass[12pt]{article}

\begin{document}

\title{On a possible new type of a T odd skewon field \\ linked to
  electromagnetism\footnote{Proceedings of the 1st Mexican Meeting on
    Mathematical and Experimental Physics, held at El Colegio
    Nacional, Mexico City, 10 to 14 September 2001.}}

\author{Friedrich W.\ Hehl\footnote{hehl@thp.uni-koeln.de} , Yuri N.\
  Obukhov\footnote{yo@thp.uni-koeln.de,
    general@elnet.msk.ru}\hskip5pt\footnote{On leave from: Department of
    Theoretical Physics, Moscow State University, 117234 Moscow,
    Russia.} , and Guillermo F.\ Rubilar\footnote{gr@thp.uni-koeln.de}\\ 
  \ \\ Institute for Theoretical Physics \\ University of Cologne \\ 
  50923 K\"oln, Germany}
  
\date{27 March 2002}\maketitle 

\begin{abstract}
  In the framework of generally covariant (pre-metric) electrodynamics
  (``charge \& flux electrodynamics''), the Maxwell equations can be
  formulated in terms of the electromagnetic excitation $H=({\cal
    D},{\cal H})$ and the field strength $F=(E,B)$. If the spacetime
  relation linking $H$ and $F$ is assumed to be {\em linear}, the
  electromagnetic properties of (vacuum) spacetime are encoded into 36
  components of the vacuum constitutive tensor density $\chi$. We
  study the propagation of electromagnetic waves and find that the
  metric of spacetime emerges eventually from the principal part
  $^{(1)}\chi$ of $\chi$ (20 independent components). In this article,
  we concentrate on the remaining skewon part $^{(2)}\chi$ (15
  components) and the axion part $^{(3)}\chi$ (1 component).  The
  skewon part, as we'll show for the first time, can be represented by
  a 2nd rank traceless tensor $\not\!\! S_i{}^j$. By means of the
  Fresnel equation, we discuss how this tensor disturbs the light
  cones.  Accordingly, this is a mechanism for violating Lorentz
  invariance and time symmetry. In contrast, the (abelian) axion part
  $^{(3)}\chi$ does {\em not} interfere with the light cones. {\em
    file mexmeet8.tex, 2002-03-27}
\end{abstract}


\section{The constitutive tensor density $\chi$ 
  of vacuum spacetime and its irreducible decomposition}

In pre-metric electrodynamics \cite{book} the axioms of electric
charge and of magnetic flux conservation manifest themselves in the
Maxwell equations for the excitation $H=({\cal D},{\cal H})$ and the
field strength $F=(E,B)$:
\begin{equation}\label{me}
dH=J, \qquad dF=0 \,.
\end{equation}
If local coordinates $x^i$ are given, with $i,j,... =0,1,2,3$, we can
decompose the excitation and field strength 2-forms into their
components according to
\begin{equation}  
H = {\frac 1 2}\,H_{ij}\,dx^i\wedge dx^j,\qquad  
F = {\frac 1 2}\,F_{ij}\,dx^i\wedge dx^j.\label{geo1}  
\end{equation} 
Then a {\em linear} spacetime relation, see \cite{PostFormal}, reads
\begin{equation}\label{HchiF}
H_{ij}={\frac 1 2}\,\kappa_{ij}{}^{kl}\,F_{kl}
=\frac{1}{4}\,{\hat \epsilon}_{ijkl}\,\chi^{klmn}\,F_{mn} \,.
\end{equation}
The constitutive tensor density $\chi$ has 36 independent components.
If we decompose it with respect to the 4-dimensional linear group into
irreducible pieces, then we find
\begin{equation}\label{chiirr1}\chi={}^{(1)}\chi + {}^{(2)}\chi + 
  {}^{(3)}\chi\,, \qquad{\rm with} \qquad 36= 20\oplus 15 \oplus
  1\end{equation} independent components, respectively. In components,
we have the following definitions for the irreducible pieces of
$\chi$:
\begin{eqnarray}\label{chiirr2}
  {}^{(2)}\chi^{ijkl}&:=&\frac{1}{2}\left( \chi^{ijkl}-
    \chi^{klij}\right)\,,\quad {}^{(3)}\chi^{ijkl}:=\chi^{[ijkl]}\,,
  \\{}^{(1)} \chi^{ijkl}&:=& \chi^{ijkl} -\,
  {}^{(2)}\chi^{ijkl}-\,{}^{(3)}\chi^{ijkl}\,.\label{chiirr3}
\end{eqnarray}
Then, by explicit substitution of the definitions, it is
straightforward to show that the irreducible piece $a$ of the
irreducible piece $b$, for $a,b=1,2,3$, with the Kronecker
$\delta^{ab}$, behaves as follows:
\begin{equation}\label{irrirr}^{(a)}\left[^{(b)}\chi^{ijkl}
  \right]=\delta^{ab}\,^{(a)}\chi^{ijkl}\,,\qquad{\rm no\; sum\; over
    \;} a\,.
\end{equation}
More explicitly, we have, e.g.,
\begin{equation}\label{sym}
    {}^{(1)} \chi^{ijkl}=\,{}^{(1)} \chi^{klij}\,,\qquad {}^{(2)}
    \chi^{ijkl}=-\,{}^{(2)} \chi^{klij}\,.
\end{equation}
A simple way to see the correctness of this irreducible decomposition
is to recall that $\chi^{ijkl}$ is antisymmetric in the first and in the
last pair of indices; then it is possible to map it to a $6\times 6$
matrix. And this matrix can be decomposed in its symmetric tracefree
piece corresponding to $^{(1)}\chi^{ijkl}$, the antisymmetric piece
$^{(2)}\chi^{ijkl}$, and its trace $^{(3)}\chi^{ijkl}$, see
(\ref{chiirr1}).

\subsection{Principal piece $^{(1)}\chi$ and light cone structure}

Let us try to get a rough picture of the physical meaning of these
different irreducible pieces. From $^{(1)}\chi^{ijkl}$ alone, if {\em
  electric/magnetic reciprocity} is assumed additionally to hold for
(\ref{HchiF}), then, up to an arbitrary conformal factor, the
Lorentzian metric of spacetime can be derived
\cite{OH99,HOR00,OFR00,GR01,nonsym32}.  One can think of this
reduction in the way that electric/magnetic reciprocity cuts the 20
components of $^{(1)}\chi^{ijkl}$ into half, that is, only 10
components for the metric are left.  Modulo the undetermined function,
we have then 9 remaining components.

These 9 components determine the {\em light cone} at each point of
spacetime. Accordingly, in $^{(1)}\chi^{ijkl}$ the light cone of
spacetime is hidden and thereby conventional Maxwell-Lorentzian vacuum
electrodymanics as well. To put it more geometrically, the first
irreducible piece of $\chi$, via electric/magnetic reciprocity, yields
the {\em conformal} structure of spacetime. In this sense, there is no
doubt that $^{(1)}\chi^{ijkl}$ is the principal part of the
constitutive tensor density $\chi$ of the vacuum.

\subsection{Abelian axion $\alpha$}

Thus one could be inclined to believe that it is best to require the
vanishing of the second and the third irreducible piece of $\chi$. But
this would appear to be premature before one has inquired into the
possible meanings of these pieces. In fact, the {\em abelian axion}
$\alpha(x)$, introduced by Ni \cite{Ni73,Ni77} in 1973, is represented
by $^{(3)}\chi^{ijkl}$.  This field is P odd (P stands for parity),
i.e., it is a pseudo- or axial-scalar in conventional language.
Experimentally, this field hasn't been found so far
\cite{Bibber,Cooper,Cheng,Stedman}. Nevertheless, as we shall see
below, the axion does not interfere with the light cone structure of
spacetime at all. Therefore, this chapter is not yet closed, the
abelian axion remains a serious option for a particle search in
experimental high energy physics.

\subsection{Skewon piece $\not\!S_i{}^j$ and dissipation} 

If two irreducible pieces of a quantity don't vanish possibly, it is
not too far-fetched to reflect on the remaining piece
$^{(2)}\chi^{ijkl}$ and on what its existence may mean. The
conventional argument for discarding $^{(2)}\chi$ runs as follows, see
Post \cite{PostFormal}.  Suppose a Lagrangian 4-form $L$ exists for
the electromagnetic field.  In general $H\sim \partial L/\partial F$.
If $H$ is assumed to be linear in $F$, as is done in (\ref{HchiF}),
then $L$ reads
\begin{equation}\label{lagrangian}
  L\sim H\wedge F=\chi\cdot F\wedge F=\,^{(1)}\chi \cdot F\wedge F+
  \,^{(3)}\chi \cdot F\wedge F\,.
\end{equation}
The term with $^{(1)}\chi$ eventually becomes the Maxwell Lagrangian,
the term with $^{(3)}\chi$ part of the axion Lagrangian. The piece
with $^{(2)}\chi$ drops out of the Lagrangian $L$ because of the
antisymmetry of $^{(2)}\chi$ according to $^{(2)}\chi^{ijkl}=
-\,^{(2)}\chi^{klij}$. Since we conventionally assume that all
information of a physical system is collected in its Lagrangian, we
reject $^{(2)}\chi^{ijkl}\ne 0$ as being unphysical.  This presents
the state of the art.

However, we would like to point out that the argument with the
Lagrangian does {\em not} forbid the existence of a non-vanishing
$^{(2)}\chi\neq 0$. It only implies that $L$ is `insensitive' to
$^{(2)}\chi$. In other words, if $^{(2)}\chi\ne 0$, then not all
information about the system is contained in the Lagrangian.

Remember that pre-metric electrodynamics is based on the conservation
laws of electric charge and magnetic flux and on an axiom about the
(kinematic) electromagnetic energy-momentum current $^{\rm
  k}\Sigma_\alpha$. No Lagrangian is needed nor assumed. But, of
course, the proto-Lagrangian $\Lambda:=-H\wedge F/2$ exists anyway and
has indeed been introduced in the context of the discussion of $^{\rm
  k}\Sigma_\alpha$. Accordingly, in pre-metric electrodynamics, even
when linearity is introduced according to (\ref{HchiF}), $\Lambda$ has
no decisive meaning --- and that $\Lambda$ does not depend on
$^{(2)}\chi$ is interesting to note but no reason for a headache.

This reminds us of a complementary property of the axion $\alpha$ or
of $^{(3)}\chi$. It features in $\Lambda$, see (\ref{lagrangian}), but
it drops out of $^{\rm k}\Sigma_\alpha$, as we will see below.  Should
we be alarmed that the axion doesn't contribute to the electromagnetic
energy-momentum current? No, not really. As has been shown by Ni, in
spite of this `insensitivity' of $^{\rm k}\Sigma_\alpha$ against
$\alpha$, one can set up a reasonable theory of the axion.

Consequently, in linear pre-metric electrodynamics, it is not alarming
that $^{(2)}\chi$ drops out from the proto-Lagrangian $\Lambda$, and
in future we will take {\em the possible existence of}
$^{(2)}\chi^{ijkl}$ {\em for granted}.

What is then the possible physical meaning of $^{(2)}\chi$? In
pre-metric electrodynamics \cite{book}, the energy-momentum current
reads
\begin{equation}
  ^{\rm k}\Sigma_\alpha :={\frac 1 2}\left[F\wedge(e_\alpha\rfloor
    H) - H\wedge(e_\alpha\rfloor
    F)\right]\,.\label{simax}
\end{equation} 
We can specify a certain vector field $\xi=\xi^\alpha e_\alpha$, with
the basis $e_\alpha$ of the tangent vector space at each point of
spacetime. Then we can transvect the energy-momentum current with
$\xi^\alpha$:
\begin{equation}\label{QQ}
  {\cal Q}:=\xi^\alpha\,^{\rm k}\Sigma_\alpha=
  \frac{1}{2}\left[F\wedge(\xi\rfloor H)- H\wedge(\xi\rfloor
    F)\right]\,.
\end{equation}
The scalar-valued 3-form $\cal Q$ is expected to be related to
conserved quantities provided we can find suitable (Killing type)
vector fields $\xi$. Therefore we determine its exterior derivative
and find after some algebra, see Rubilar \cite{Dr.Guillermo},
\begin{equation}\label{dQQ}
  d{\cal Q}=(\xi\rfloor F)\wedge J+\frac{1}{2}\left( F\wedge {\cal
      L}_\xi H-H\wedge{\cal L}_\xi F \right)\,,
\end{equation} 
or, in holonomic components, with ${\cal Q}^i:=
\epsilon^{ijkl}Q_{jkl}/6$, $\,{\cal J}^i:= \epsilon^{ijkl}J_{jkl}/6$,
and ${\cal H}^{ij} :=\epsilon^{ijkl}H_{kl}/2$,
\begin{equation}\label{diqq}
  \partial_i {\cal Q}^i=\xi^kF_{kl}\,{\cal J}^l+\frac{1}{4}\left(
    F_{kl}\,{\cal L}_\xi {\cal H}^{kl} -{\cal H}^{kl}\, {\cal L}_\xi
    F_{kl}\right)\,.
\end{equation}
Here ${\cal L}_\xi$ denotes the Lie derivative along $\xi$.  Now we
substitute the {\em linear relation} (\ref{HchiF}), or ${\cal
  H}^{kl}=\chi^{klmn}F_{mn}/2$, and find
\begin{equation}\label{diqqlin}
  \partial_i {\cal Q}^i=\xi^kF_{kl}\,{\cal J}^l+\frac{1}{8}\left[
    F_{kl}\,{\cal L}_\xi (\chi^{klmn}F_{mn}) -\chi^{klmn}F_{mn}\,
    {\cal L}_\xi F_{kl}\right]\,.
\end{equation}
We apply the Leibniz rule of the Lie derivative and rearrange a bit:
\begin{equation}\label{diqqlin1}
  \partial_i {\cal Q}^i=\xi^kF_{kl}\,{\cal J}^l+\frac{1}{8}\left[
    ({\cal L}_\xi \chi^{ijkl})\,F_{ij}F_{kl}+(\chi^{ijkl}-
    \chi^{klij})\,F_{ij}\, {\cal L}_\xi F_{kl}\right]\,.
\end{equation}
We substitute the irreducible pieces of $\chi^{ijkl}$. Then we have
\begin{equation}\label{diqqlin2}
  \partial_i {\cal Q}^i=\xi^kF_{kl}\,{\cal J}^l+\frac{1}{8}\, {\cal
    L}_\xi \left(^{(1)}\chi^{ijkl}+\,^{(3)}\chi^{ijkl}
  \right)\,F_{ij}F_{kl}+ \frac{1}{4}\,^{(2)}\chi^{ijkl} \,F_{ij}\,
  {\cal L}_\xi F_{kl}\,.
\end{equation}

If $^{(1)}\chi$ and $^{(3)}\chi$ carry a reasonable symmetry, namely $
{\cal L}_\xi\,^{(1)}\chi={\cal L}_\xi\,^{(2)}\chi$ $=0$, and are thus
well-behaved, then, in vacuum, i.e., for ${\cal J}^i=0$, we have
non-conservation of energy, for example, because of the offending term
$^{(2)}\chi\, F\dot{ F}$. Here the dot symbolizes the `time'
derivative along $\xi$.

In any case we see that $^{(2)}\chi$ induces a {\em dissipative} term
with first `time' derivative. This is what we might have expected
since dissipative phenomena in general cannot be described in a
Lagrangian framework.

It is then our hypothesis that $^{(2)}\chi$ can represent fields which
are {\em odd under T transformations}.  Of course, we must investigate
how these {\sl skewons}, as we may call them in a preliminary way,
disturb the light cone and whether there is perhaps a viable subclass
of the skewons.

\section{The skewon field $\not\! S_i{}^j$}

The {\em skewon} piece of the constitutive tensor density
$\chi^{ijkl}$ is defined in (\ref{chiirr2})$_1$ Therefrom we can read
off the algebraic symmetries
\begin{equation}\label{2chisym1}
  {}^{(2)}\chi^{ijkl}=- {}^{(2)}\chi^{klij}\,,\qquad
  {}^{(2)}\chi^{[ijkl]}=0\,.
\end{equation}
>From $\chi^{ijkl}$, the skewon piece inherits the antisymmetry in the
first and the second pair of indices:
\begin{equation}\label{2chisym2}
 {}^{(2)}\chi^{(ij)kl}=0\,,\qquad {}^{(2)}\chi^{ij(kl)}=0\,.
\end{equation}

Thus $^{(2)}\chi^{ijkl}$ can be mapped to an antisymmetric (or {\em
  skew\/}sym\-metric) $6\times 6$ matrix. For this reason we called it
the skewon piece of $\chi^{ijkl}$. Since this matrix has 15
independent components, we expect that it is equivalent to a
2nd rank tensor in 4 dimensions (16 components) with vanishing trace
(1 component).  Accordingly, we define the skewon field by
\begin{equation}\label{skewon}
  S_i{}^j:=\frac{1}{4}\,\hat{\epsilon}_{iklm}{}^{(2)}\chi^{klmj}\,.
\end{equation}
Because of (\ref{2chisym1})$_2$, its trace vanishes, indeed,
\begin{equation}
  S_n{}^n=\frac{1}{4}\,\hat{\epsilon}_{nklm}{}^{(2)}
  \chi^{[klmn]}=0\,.
\end{equation}
If we define the tracefree part of $S_i{}^j$ by 
\begin{equation}\label{dtrace}
 \not\!S_i{}^j:= S_i^{\ j}-\frac{1}{4}\,S_k{}^k\delta_i^j\,,
\end{equation}
then for our $S_i{}^j$, we have $S_i{}^j=\not\! S_i{}^j$.

Let us invert (\ref{skewon}). We multiply by $\epsilon^{inpq}$ and find
\begin{equation}
  \epsilon^{inpq}\,S_i{}^j:=\frac{1}{4}\,\epsilon^{inpq}\,
  \hat{\epsilon}_{iklm}{}^{(2)}\chi^{klmj}=\frac{1}{4}\,\delta^{npq}
  _{klm}\,{}^{(2)}\chi^{[klm]j}\,
\end{equation}or
\begin{equation}\label{2chiins}
{}^{(2)}\chi^{[ijk]l}=-\frac{2}{3}\,\epsilon^{ijkm}\,S_m{}^l\,.
\end{equation}
We expand the bracket:
\begin{equation}
  {}^{(2)}\chi^{ijkl}+{}^{(2)}\chi^{jkil}
  +{}^{(2)}\chi^{kijl} =-2\,\epsilon^{ijkm}\,S_m{}^l\,.
\end{equation}
The second term on the left hand side of this equation, by means of the
symmetries (\ref{2chisym1})$_1$ and (\ref{2chisym2})$_1$, can be
rewritten as
$^{(2)}\chi^{jkil}=-{}^{(2)}\chi^{iljk}={}^{(2)}\chi^{lijk}$. Thus,
\begin{equation}
  {}^{(2)}\chi^{ijkl}+2{}^{(2)}\chi^{(k|ij|l)}
  =-2\,\epsilon^{ijkm}\,S_m{}^l
\end{equation}or, because of (\ref{2chisym2})$_2$,
\begin{equation}\label{semiresult}
  {}^{(2)}\chi^{ijkl}
  =2\,\epsilon^{ijm[k}\,S_m{}^{l]}\,.
\end{equation}
In order to make the symmetry (\ref{2chisym1})$_1$ manifest, we rename
the indices
\begin{equation}\label{semiresult'}
  {}^{(2)}\chi^{klij}
  =2\,\epsilon^{klm[i}\,S_m{}^{j]}
\end{equation}and subtract (\ref{semiresult'}) from (\ref{semiresult}).
This yields the final result
\begin{equation}\label{result}
  {}^{(2)}\chi^{ijkl} =\epsilon^{ijm[k}\,S_m{}^{l]}
    -\epsilon^{klm[i}\,S_m{}^{j]}\,.
\end{equation}
For (\ref{result}), all the symmetries (\ref{2chisym1}) and
(\ref{2chisym2}) can be verified straightforwardly. In (\ref{skewon}),
we chose the conventional factor as $1/4$ in order to find in
(\ref{result}) a formula free of inconvenient factors.

\section{Decomposing the ``dual'' constitutive tensor density 
  $\kappa_{ij}{}^{kl}$ and recovering the skewon field}

Let us recall that the starting point for the discussion of the
constitutive (spacetime) relation is the $\kappa$-map. Namely, we have
the tensor density $\kappa_{ij}{}^{kl}$ with 36 components, see
\cite{book} Eq.(D.1.11). One can decompose this object into its
irreducible pieces. Obviously, contraction is the only tool for such a
decomposition. Following Post \cite{Postmap}, we can define the
contacted tensor of type (1,1),
\begin{equation}
\kappa_i{}^k := \kappa_{il}{}^{kl}\,,
\end{equation}with 16 independent components. 
The second contraction yields the pseudo-scalar function
\begin{equation}
\kappa := \kappa_k{}^k = \kappa_{kl}{}^{kl}\,.
\end{equation}
The traceless piece
\begin{equation}
\not\!\kappa_i{}^k := \kappa_i{}^k - {\frac 1 4}\,\kappa\,\delta_i^k
\end{equation}has 15 independent components. 
These pieces can now be subtracted from the original constitutive
tensor. Then,
\begin{eqnarray}
  \kappa_{ij}{}^{kl} &=& {}^{(1)}\kappa_{ij}{}^{kl} +
  {}^{(2)}\kappa_{ij}{}^{kl} + {}^{(3)}\kappa_{ij}{}^{kl} \\ &=&
  {}^{(1)}\kappa_{ij}{}^{kl} +
  2\!\not\!\kappa_{[i}{}^{[k}\,\delta_{j]}^{l]} + {\frac 1
    6}\,\kappa\,\delta_{[i}^k\delta_{j]}^l.\label{kap-dec}
\end{eqnarray}
By construction, ${}^{(1)}\kappa_{ij}{}^{kl}$ is the totally traceless
part of the constitutive map:
\begin{equation}
{}^{(1)}\kappa_{il}{}^{kl} = 0.\label{notrace}
\end{equation}
Thus, we split $\kappa$ according to $36 = 20 + 15 + 1$, and the (2,2)
tensor ${}^{(1)}\kappa_{ij}{}^{kl}$ is subject to the 16 constraints
(\ref{notrace}) and carries $20 = 36 -16$ components.
%

Now we are prepared to proceed with the analysis of the
$\chi$-picture.  By definition, we have
\begin{equation}
\chi^{ijkl} := {\frac 1 2}\,\epsilon^{ijmn}\,\kappa_{mn}{}^{kl}.
\end{equation}
Substituting here the decomposition (\ref{kap-dec}), we find
\begin{equation}
\chi^{ijkl} = {}^{(1)}\chi^{ijkl} + {}^{(2)}\chi^{ijkl}
+ {}^{(3)}\chi^{ijkl}.\label{chi-dec}
\end{equation}
In correspondence with (\ref{kap-dec}), we have the irreducible
pieces:
\begin{eqnarray}
{}^{(1)}\chi^{ijkl} &=& {\frac 1 2}\,\epsilon^{ijmn}\,\,{}^{(1)}
\kappa_{mn}{}^{kl},\\ 
{}^{(2)}\chi^{ijkl} &=& {\frac 1 2}\,\epsilon^{ijmn}\,\,{}^{(2)}
\kappa_{mn}{}^{kl} = -\,\epsilon^{ijm[k}\!\not\!\kappa_m{}^{l]},\\
{}^{(3)}\chi^{ijkl} &=& {\frac 1 2}\,\epsilon^{ijmn}\,\,{}^{(3)}
\kappa_{mn}{}^{kl} = {\frac 1 {12}}\,\epsilon^{ijkl}\,\kappa. 
\end{eqnarray}

Let us identify the skewon and the axion fields by
\begin{equation}
S_i{}^j = -\,{\frac 1 2}\!\not\!\kappa_i{}^j,\qquad
\alpha = {\frac 1 {12}}\,\kappa.
\end{equation}
Using the S-identity (\ref{Sident}), we have
\begin{equation}
  {}^{(2)}\chi^{ijkl} = 2\,\epsilon^{ijm[k}\,S_m{}^{l]} =
  -\,2\,\epsilon^{klm[i}\,S_m{}^{j]}
\end{equation}
or
\begin{equation}
  {}^{(2)}\chi^{ijkl} = \epsilon^{ijm[k}\,S_m{}^{l]} -
  \epsilon^{klm[i}\,S_m{}^{j]}.
\end{equation}
Thus, the S-identity (\ref{Sident}) guarantees the {\it skew-}symmetry
of $^{(2)}\chi$ under exchange of the first with the second index
pair:
\begin{equation}
{}^{(2)}\chi^{ijkl} = -\,{}^{(2)}\chi^{klij}.
\end{equation}
On the other hand, the K-identity (\ref{Kident}) provides the {\it
  symmetry} of the $^{(1)}\chi$:
\begin{equation}
{}^{(1)}\chi^{ijkl} = {}^{(1)}\chi^{klij}.
\end{equation}
This holds true because of the tracelessness property (\ref{notrace}).

It is thus very satisfactory to find the one-to-one correspondence of
the irreducible decomposition (\ref{kap-dec}) of $\kappa_{ij}{}^{kl}$
[based on the trace extraction] and the irreducible decomposition
(\ref{chi-dec}) of $\chi^{ijkl}$ [based on the separation into
symmetric and skew-symmetric parts].

\section{The skewon field as $6\times 6$ matrix}

It is convenient to put the $S_i{}^j$ also into the conventional
$6\times 6$ matrix since this provides the interpretation of the
spacetime relation in terms of the 3-dimensional electromagnetic field
${\cal D},{\cal H}, B,E$. Therefore we compute the $3\times 3$
matrices with the help of (\ref{result}) in a fairly messy but
straightforward way, see \cite{Dr.Guillermo}:
\begin{eqnarray}\label{A-matrix}
  ^{(2)}{\cal A}^{ba}&:=&{}^{(2)}\chi^{0a0b}\quad
  =\epsilon^{abc} S_c{}^0\,,\\ 
\label{B-matrix}
^{(2)}{\cal B}_{ba}&:=&\frac{1}{4}\,\hat\epsilon_{acd}\, \hat\epsilon_{bef}
\,{}^{(2)}\chi^{cdef}\quad=-\hat\epsilon_{abc}\, S_0{}^c\,,\\ 
\label{C-matrix}
^{(2)}{\cal C}^a{}_b& :=&\frac{1}{2}\,\hat\epsilon_{bcd}\, {}^{(2)}
\chi^{cd0a}\quad =-S_b{}^a+\delta_b^a\,S_c{}^c\,,\\ 
\label{D-matrix}
^{(2)}{\cal D}_a{}^b&:=&\frac{1}{2}\,\hat\epsilon_{acd}
\,{}^{(2)}\chi^{0bcd} \quad = S_a{}^b-\delta_a^b\,S_c{}^c\,.
\end{eqnarray}

Quite generally, we have
\begin{equation}\chi^{IJ}= \left( \begin{array}{cc} {\cal B}_{ab}&
      {\cal D}_a{}^b \\ {\cal C}^a{}_b & {\cal A}^{ab} \end{array}\right)\,.
\end{equation} For the skewon piece\footnote{For future reference 
  we display here also the principal and the axion pieces as $6\times
  6$ matrices, respectively, namely
\begin{equation}
  ^{(1)}\chi^{IJ}= \left( \begin{array}{cc} {\cal B}_{(ab)}&
      \hspace{-0.4truecm}\frac{1}{2}\left(\not\!\! {\cal
          D}_a{}^b + \not\!{\cal C}^b{}_a\right) \\ 
      \frac{1}{2}\left(\not\! {\cal C}^a{}_b + \not\!\!{\cal D}_b{}^a\right)
      &\hspace{-0.4truecm} {\cal A}^{(ab)}
    \end{array}\right) = \left( \begin{array}
      {cc} ^{(1)} {\cal B}_{ab}&\hspace{-0.25truecm} ^{(1)} {\cal
        D}_a{}^b \\ ^{(1)}{\cal C}^a{}_b &\hspace{-0.25truecm} ^{(1)}
      {\cal A}^{ab}
    \end{array}\right)\,,\nonumber\end{equation}
here we used the notation $\not\!\!M_a{}^b :=
M_a{}^b-M_c{}^c\,\delta_a^b/3$, and 
\begin{equation}
  ^{(3)}\chi^{IJ}= \frac{1}{6}\left({\cal C}^c{}_c+{\cal D}_c{}^c
  \right)\left(\begin{array}{cc}0_3 & 1_3\\ 1_3&
      0_3\end{array}\right)= \frac{1}{6}\left({\cal C}^c{}_c 
     + {\cal D}_c{}^c \right)\,\epsilon^{IJ}\,.
\end{equation}}, we have specifically
\begin{equation}
  ^{(2)}\chi^{IJ}= \left( \begin{array}{cc} {\cal
        B}_{[ab]}&\frac{1}{2}\left( {\cal D}_a{}^b-{\cal
          C}^b{}_a\right) \\ \frac{1}{2}\left( {\cal C}^a{}_b-{\cal
          D}_b{}^a\right) & {\cal A}^{[ab]}
    \end{array}\right) = \left( \begin{array}
      {cc} ^{(2)} {\cal B}_{ab}&\hspace{-0.25truecm} ^{(2)} {\cal
        D}_a{}^b \\ ^{(2)}{\cal C}^a{}_b &\hspace{-0.25truecm} ^{(2)}
      {\cal A}^{ab}
    \end{array}\right)\nonumber
\end{equation}
\begin{equation}\label{2matrix} 
  =\left(
\begin{array}{cc}-\,\hat\epsilon_{abc} S_0{}^c&+S_a{}^b-\delta^b_a\,S_c{}^c\\
  -S_b{}^a+\delta_b^a\,S_c{}^c&\epsilon^{abc}\,S_c{}^0\end{array}
\right)\,.
\end{equation} 
The rest is done easily.  We substitute (\ref{2matrix}) into the
spacetime relation
\begin{equation} 
  \left(\begin{array}{c} {\cal H}_a \\ {\cal {D}}^a\end{array}\right)
  = \left(\begin{array}{cc} {{\cal C}}^{b}{}_a &
      {{\cal B}}_{ba} \\ {{\cal A}}^{ba}&
      {{D}}_{b}{}^a \end{array}\right)
  \left(\begin{array}{c} -E_b\\ B^b\end{array}\right)\,\label{CR'}
\end{equation}
and find quite generally for the skewon contributions of the 3D
excitations,
\begin{eqnarray}\label{Dfield}
  ^{(2)}{\cal D}^a&=& -\epsilon^{abc}S_c{}^0E_b
  +(-\delta_b^a\,S_c{}^c + S_b{}^a)B^b\,,\\ ^{(2)}{\cal
    H}_a&=&\;(-\delta_a^b \,S_c{}^c + S_a{}^b)E_b-
  \hat\epsilon_{abc}\,S_0{}^cB^b\,.\label{Hfield}
\end{eqnarray} 
The diagonal terms of the 3D tensor $(-\delta_a^b \,S_c{}^c +
  S_a{}^b)$ are of the type as those postulated
by Nieves and Pal \cite{NP94} for describing a ``...third
electromagnetic constant of an isotropic medium''. However, our
``medium'' is spacetime, i.e., the vacuum itself.

We stress that for the derivation of (\ref{Dfield}) and (\ref{Hfield})
we neither specialized the skewon field $S_i{}^j$ nor did we apply any
metric distilled from $^{(1)}\chi^{ijkl}$. Therefore the 1+3
decompositions in (\ref{Dfield}) and (\ref{Hfield}) are generally
valid for {\em any linear} spacetime relation.

\section{Spatially isotropic skewon field and the\\
 ansatz of Nieves and Pal}

If we specialize first to 3-dimensional {\it isotropy}, then we have,
with the 3D pseudo-scalar function $s=s(x)$,
\begin{equation}\label{isotropy}
  S_a{}^b = {\frac s 2}\,\delta_a^b\,,\qquad S_0{}^a = 0\,,\qquad
  S_b{}^0 =0\,.
\end{equation}
It was possible to formulate isotropy for the skewon piece {\it
  without} taking recourse to a metric tensor since $S_i{}^j$ is a
mixed variant tensor of 2nd rank. Thus,
\begin{equation}
  S_i{}^j = \frac{s}{2}\left(\begin{array}{rccc} -3 & 0 & 0 & 0 \\ 0 &
      1 & 0 & 0 \\ 0 & 0 & 1 & 0 \\ 0 & 0 & 0 & 1
    \end{array}\right)
\end{equation} and
\begin{equation}
  -\delta_a^b\,S_c{}^c + S_a{}^b = -s\,\delta_a^b\,.
\end{equation}
Consequently, equations (\ref{Dfield}) and (\ref{Hfield}) become
\begin{equation}
  ^{(2)}{\cal D}^a=-s\,B^a\,,\qquad^{(2)}{\cal H}_a=
  -s\,E_a\,,\label{DH-new1a}
\end{equation}
exactly what Nieves and Pal had postulated and discussed subsequently
\cite{NP94}. Accordingly, the spacetime relations (\ref{Dfield}) and
(\ref{Hfield}) are {\it anisotropic} generalizations of the Nieves and
Pal ansatz. The off diagonal terms with $S_0{}^a$ and $S_b{}^0$ lead,
respectively, to magnetic and electric Faraday type of effects of the
spacetime under consideration, i.e., these terms rotate the
polarization of a wave propagating in such a spacetime.

\section{On the four electromagnetic constants for 
  vacuum spacetime with spatial isotropy}

Since the axion field also contributes to the spacetime relation of
the type (\ref{DH-new1a}), we are going to determine it. We have
\begin{equation}
  ^{(3)}\chi^{ijkl}:=\alpha\,\epsilon^{ijkl}\,,\quad
  \alpha=\frac{1}{4!}\,\hat{\epsilon}_{ijkl}\, ^{(3)}\chi^{ijkl}\,,
\end{equation}
\begin{equation}
  ^{(3)}\chi^{IJ}=\alpha\left(\begin{array}{cc}0_3 & 1_3\\ 1_3&
      0_3\end{array}\right)=\alpha\,\epsilon^{IJ}\,.
\end{equation}
Because of (\ref{CR'}), we find
\begin{eqnarray} 
  ^{(3)} {\cal D}^a &=&+\alpha\,B^a\,, \\ ^{(3)} {\cal H}_a
  &=&-\alpha\,E_a\,.
\end{eqnarray}
If we take (\ref{principal-metric}) from below in Cartesian
coordinates, i.e., we have a Lorentz metric
$o_{ij}\stackrel{\ast}{=}(c^2,-1,-1,-1)$ with
$o_{ab}\stackrel{\ast}{=}-\delta_{ab}$, then the spacetime relation
becomes
\begin{eqnarray}\label{resultx}
  {\cal D}^a &=&\;\, \varepsilon_0\,\delta^{ab} E_b\; +\left(-s+
    \alpha\right)B^a\,, \\ {\cal H}_a &=& \left(-s-
    \alpha\right)E_a+\mu_0^{-1}\,\delta_{ab}\, B^b\,.
\end{eqnarray}
In the special case when skewon and axion become {\it constant}
fields, we can say that we found 4 electromagnetic constants for a
spacetime (vacuum) with spatial isotropy: The electric constant
$\varepsilon_0$, the magnetic constant $\mu_0$, the isotropic part $s$
of the skewon $S_a{}^b$, and the axion $\alpha$.

\section{How does the skewon field affect light  propagation 
\cite{nonsym32,Dr.Guillermo}?}

For any linear spacetime relation, the Fresnel equation can be written
as
\begin{equation} \label{Fresnel}
{\cal G}^{ijkl}q_i q_j q_k q_l = 0 \,,
\end{equation}where $q_i$ is the wave covector.
The fourth order tensor density of weight $+1$, the {\it Fresnel
  tensor}, as we may call it, is defined by
\begin{equation}\label{G4}
 {\cal G}^{ijkl}:=\frac{1}{4!}\,\hat{\epsilon}_{mnpq}\,
 \hat{\epsilon}_{rstu}\,\chi^{\,mnr(i}\, \chi^{\,j|ps|k}\, \chi^{\,l)qtu }\,.
\end{equation}

First, we recall \cite{nonsym32} that the Fresnel equation is
independent of the axion piece ${}^{(3)}\chi$ of the constitutive
tensor:
\begin{equation}\label{propg1}
 {\cal G}^{ijkl}(\chi)= {\cal G}^{ijkl}({}^{(1)}\chi+{}^{(2)}\chi) .
\end{equation}
Thus, for arbitrary $^{(1)}\chi$ and $^{(2)}\chi$, we have
\begin{equation} \label{propg2}
  {\cal G}^{ijkl}(^{(3)}\chi)=0 .
\end{equation}
Furthermore, due to the skewsymmetry (\ref{2chisym1})$_1$ of
${}^{(2)}\chi$, we have
\begin{equation}\label{propg3}
  {\cal G}^{ijkl}({}^{(2)}\chi)= 0 .
\end{equation}
By explicit calculations, we used computer algebra for it, we find
\begin{equation} \label{propg4}
  {\cal G}^{ijkl}({}^{(2)}\chi+{}^{(3)}\chi)=0\,.
\end{equation}
This identity is non-trivial since $\cal G$ depends cubically on the
constitutive tensor $\chi$. The identity (\ref{propg4}) shows that the
symmetric piece $^{(1)}\chi$ is indispensable in order to obtain well
behaved wave properties: If $^{(1)}\chi=0$, the Fresnel equation is
trivially satisfied and thus no light cone structure could be induced.

Furthermore, since in general
\begin{equation} \label{propg5}
  {\cal G}^{ijkl}({}^{(1)}\chi+ {}^{(2)}\chi)\neq {\cal
    G}^{ijkl}({}^{(1)}\chi) ,
\end{equation}
the skewon field {\it does} influences the Fresnel equation, and
therefore, eventually the light cone structure. An example of this
general result can be found in the asymmetric constitutive tensor
studied by Nieves and Pal \cite{NP89,NP94}.  Actually, after some
algebra one finds
\begin{eqnarray} \label{propg6}
  {\cal G}^{ijkl}({}^{(1)}\chi+{}^{(2)}\chi)&=&{\cal
    G}^{ijkl}({}^{(1)}\chi) \nonumber \\ &+& 
  \frac{2}{4!}\,\hat{\epsilon}_{mnpq}\,
  \hat{\epsilon}_{rstu}\,{}^{(1)}\chi^{\,mnr(i}\,
  {}^{(2)}\chi^{\,j|ps|k}\, {}^{(2)}\chi^{\,l)qtu } \nonumber \\ &+&
  \frac{1}{4!}\,\hat{\epsilon}_{mnpq}\,
  \hat{\epsilon}_{rstu}\,{}^{(2)}\chi^{\,mnr(i}\,
  {}^{(1)}\chi^{\,j|ps|k}\, {}^{(2)}\chi^{\,l)qtu }
\end{eqnarray}
or, in a (more of less) obvious notation, see the definition
(\ref{G4}),
\begin{eqnarray} \label{propg7}
  {\cal G}^{ijkl}(\chi,\chi,\chi)&=& {\cal
    G}^{ijkl}({}^{(1)}\chi,{}^{(1)}\chi,{}^{(1)}\chi) + 2\, {\cal
    G}^{ijkl}({}^{(1)}\chi,{}^{(2)}\chi,{}^{(2)}\chi) \nonumber \\ &&
  + {\cal G}^{ijkl}({}^{(2)}\chi,{}^{(1)}\chi,{}^{(2)}\chi)\, .
\end{eqnarray}
The other terms vanish due to the symmetry properties of each
irreducible piece.

Take now (\ref{propg6}) and substitute the parametrization of
${}^{(2)}\chi$ in terms of $S_i^{\ j}$, see (\ref{skewon}).  After
some lengthy but straightforward algebra, one finds that the two last
contributions to the right hand side of (\ref{propg6}) are actually
equal, namely
\begin{eqnarray}
 {\cal G}^{ijkl}({}^{(1)}\chi,{}^{(2)}\chi,{}^{(2)}\chi)&=&
 {\cal G}^{ijkl}({}^{(2)}\chi,{}^{(1)}\chi,{}^{(2)}\chi) \\
 &=& \frac{1}{3}\,{}^{(1)}
 \chi^{\,m(i|n|j}S_m^{\ k} S_n^{\ l)}.
\end{eqnarray}
Therefore, the final result reads
\begin{equation} \label{propg8}
  {\cal G}^{ijkl}(\chi) = {\cal G}^{ijkl}({}^{(1)}\chi + {}^{(2)}\chi)
  = {\cal G}^{ijkl}({}^{(1)}\chi) + {}^{(1)}\chi^{\,m(i|n|j}S_m^{\ k}
  S_n^{\ l)}\,,
\end{equation}
a very simple expression, indeed.

For the particular case of Nieves and Pal, we will use the ansatz
\begin{equation}\label{split}
  S_i{}^j=\not\!
  S_i{}^j=\omega_i\,v^j-\frac{1}{4}\,\omega_k\,v^k\,\delta_i^j
\end{equation} and assume additionally the existence of a metric
$g_{ij}$ (resulting from $^{(1)}\chi$!) for raising and lowering
indices: $\omega_i=g_{il}\,v^l$.  This determines $^{(2)}\chi$ via
(\ref{result}). Furthermore, we assume for the principal part the
usual metric dependent expression for the vacuum in a Riemannian
spacetime, namely
\begin{equation}\label{principal-metric}
  ^{(1)}\chi^{ijkl}=2
  \sqrt{\frac{\varepsilon_0}{\mu_0}}\sqrt{-g}\,g^{i[k}g^{l]j}\,.
\end{equation}
Because of (\ref{propg1}), the axion piece is not required.
Accordingly, we substitute $^{(1)}\chi$ and $^{(2)}\chi$ into
(\ref{G4}).  Then,
\begin{eqnarray}
  {\cal G}^{ijkl}q_iq_jq_kq_l
  &=&-\sqrt{\frac{\varepsilon_0}{\mu_0}}\sqrt{-g} \times \nonumber \\ 
  &&\hspace{-0.5cm} \left[\frac{\varepsilon_0}{\mu_0} (q\cdot q)^2
    -(q\cdot q)(v\cdot v)(v\cdot q)^2+(v\cdot
    q)^4\right]=0\,.\label{NPFres}
\end{eqnarray}
We now use $q_i\stackrel{*}{=}(\omega,-\vec{k})$,
$\,v^i\stackrel{*}{=} (v,0,0,0)$, $\,g_{ij}= o_{ij}\stackrel{*}{=}
(c^2,-1,-1,-1)$, and $\,c^2=\frac{1}{\varepsilon_0\mu_0}$. Thus,
\begin{equation}\label{NPFres'}
  -\frac{1}{\varepsilon_0^3}\, {\cal
    G}^{ijkl}q_iq_jq_kq_l\,\stackrel{*}{=}\,\left[
  \omega^2-(c\vec{k})^2\right]^2
  +\left[\frac{c{v}^2}{\varepsilon_0}\,\omega\, c\vec{k}\right]^2=0\, .
\end{equation}
This equation describes how the skewon piece, via $v$, affects light
propagation. In general, for $v\neq 0$, the Fresnel equation will
have complex solutions. This is again a manifestation of the
dispersive properties described by the skewon piece ${}^{(2)}\chi$ of
the constitutive tensor. Modulo different conventions, our result
(\ref{NPFres'}) agrees with that of Nieves and Pal \cite{NP94}
Eq.(5.7).

\section{Discussion}

The skewon and the axion part of $\chi$ are explicitly known:
\begin{eqnarray}
  {}^{(2)}\chi^{ijkl}&=&\epsilon^{ijm[k}\,S_m{}^{l]}
  -\epsilon^{klm[i}\,S_m{}^{j]} \,, \\ 
  ^{(3)}\chi^{ijkl}&=&\alpha\,\epsilon ^{ijkl}\,.
\end{eqnarray}
Accordingly, we found a traceless 2nd rank tensor field
$S_i{}^j=\not\!S_i{}^j$ and a pseudo-scalar $\alpha$. For the principal
part $^{(1)}\chi$ with its 20 independent components things are more
difficult. 

We can tentatively assume
\begin{equation}\label{1ans}
  ^{(1)}\chi^{ijkl}\sim g^{[i|[k}\,h^{l]|j]} + g^{[k|[i}\,h^{j]|l]} +
  \epsilon^{ijm[k}\,a_m{}^{l]} +\epsilon^{klm[i}\,a_m{}^{j]}\,,
\end{equation}
with two symmetric $g^{ij}=g^{ji},\,h^{ij}=h^{ji}$ and a traceless
tensor $a_i{}^j$, with $a_k{}^k=0$. However, a second look
convinces us that in (\ref{1ans}) the two last terms (with $a$) should
be deleted. The reason is the S-identity, see Appendix. In view of
(\ref{Sident}), a traceless (1,1) tensor can only contribute to the
$^{(2)}\chi$, not to $^{(1)}\chi$.  So, the structure of $^{(1)}\chi$
is most probably determined only by the two symmetric tensors $g^{ij}$
and $h^{ij}$ with $10+10$ independent components.

Furthermore, it is clear that the two terms on the r.h.s. of
(\ref{1ans}) are equal: $g^{[i|[k}\,h^{l]|j]} = g^{[k|[i}\,h^{j]|l]}$.
Indeed:
\begin{eqnarray}
  4g^{[i|[k}\,h^{l]|j]} &=& g^{ik}h^{lj} - g^{il}h^{kj} - g^{jk}h^{li}
  + g^{jl}h^{ki} =  4g^{[k|[i}\,h^{j]|l]}\\ &=& g^{ki}h^{jl} -
  g^{kj}h^{il} - g^{li}h^{jk} + g^{lj}h^{ik}.
\end{eqnarray}
Since both tensors are symmetric, $g^{ij}=g^{ji}$ and $h^{ij}=h^{ji}$,
the two expressions are equivalent. Thus only one term is left over:
\begin{equation}\label{1ans'}
  ^{(1)}\chi^{ijkl}\sim g^{[i|[k}\,h^{l]|j]}
\end{equation}

By the way, if we turn to the $\kappa$-representation, we have:
\begin{equation}
  ^{(1)}\kappa_{ij}{}^{kl} \sim \hat\epsilon_{ijmn}\,g^{m[k}h^{l]n}\,.
\end{equation}
For the symmetric $g^{ij}=g^{ji}$ and $h^{ij}=h^{ji}$, a contraction
is automatically zero, $^{(1)}\kappa_{il}{}^{kl} \equiv 0$. This
means that such a term belongs indeed to the first irreducible part, in
accordance with Sec.3. Such a structure looks very much as the 
most general parametrization of the first irreducible part, but
a proof is still not available.

%
Summarizing: it seems that the general structure of the first irreducible
part reads
\begin{equation}
^{(1)}\chi^{ijkl} \sim g^{[i|[k}\,h^{l]|j]}.
\end{equation}
It would be desirable to find a corresponding proof.\medskip

\subsection*{Acknowledgments}
  G.F.R.\ would like to thank the German Academic Exchange Service
  (DAAD) for financial support. We are grateful to Alfredo
  Mac\'{\i}as, UAM-Iztapalapa, for inviting us to contribute to the
  Mexican Physics Meeting quoted above. This project has been partly
  supported by CONACyT Grants: 28339E, 32138E, by a FOMES Grant:
  P/FOMES 98--35--15, and by the joint German--Mexican project CONACyT
  --- DFG: E130--655 --- 444 MEX 100.

\appendix
\section{Two identities}

We can rederive our results (\ref{skewon}) and (\ref{result}) in an
alternative way in order to get more insight in the relevant
structure. In 4D, any object with five completely antisymmetrized
indices is zero, $Z^{[ijmkl]}_{...}\equiv 0$. When 4 of these 5
indices belong to the Levi-Civita symbol, we have the identity:
\begin{equation}
\epsilon^{ijmk}\,Z^l_{...} \equiv \epsilon^{ljmk}\,Z^i_{...} +
\epsilon^{ilmk}\,Z^j_{...} + \epsilon^{ijlk}\,Z^m_{...} +
\epsilon^{ijml}\,Z^k_{...}. \label{Zident}
\end{equation}
Applying this for the case when $Z = S_m{}^l$, we find the identity
\begin{equation}
\epsilon^{ijmk}\,S_m{}^l \equiv \epsilon^{ljmk}\,S_m{}^i +
\epsilon^{ilmk}\,S_m{}^j + \epsilon^{ijlk}\,S_m{}^m +
\epsilon^{ijml}\,S_m{}^k.
\end{equation}
Suppose that $S_i{}^j$ is a {\it traceless} tensor, i.e.\ $S_m{}^m =0$.
Then a simple rearrangement of the terms in the above identity yields:
\begin{equation}
  \epsilon^{ijmk}\,S_m{}^l - \epsilon^{ijml}\,S_m{}^k \equiv -
  \epsilon^{klmi}\,S_m{}^j + \epsilon^{klmj}\,S_m{}^i.
\end{equation}

Summarizing, we proved the {\bf S-identity}: {\it Every (1,1) tensor
  $S_i{}^j$ which is traceless, $S_k{}^k=0$, has the property}
\begin{equation}\label{Sident}
  \epsilon^{ijm[k}\,S_m{}^{l]} \equiv
  -\,\epsilon^{klm[i}\,S_m{}^{j]}\,.
\end{equation}
This identity always holds true in four dimensions, just because of
the properties of the Levi-Civita symbol. Eq.(\ref{Sident}) can also
be found by adding (\ref{semiresult}) and (\ref{semiresult'}).  The
S-identity underlies the possibility to express $^{(2)}\chi$ in terms
of the skewon tensor.

Let us now demonstrate another identity which holds for a (2,2)
tensor $K_{ij}{}^{kl}$. Consider the contraction
\begin{equation}
\epsilon^{ijmn}\,K_{mn}{}^{[kl]}
= {\frac 1 2}\left(\epsilon^{ijmn}\,K_{mn}{}^{kl} -
\epsilon^{ijmn}\,K_{mn}{}^{lk}\right).\label{Kid1}
\end{equation}
Apply the identity (\ref{Zident}) to the indices $[ijmnk]$ in the first
term on the r.h.s.\ and to the indices $[ijmnl]$ in the second term:
\begin{eqnarray}
\epsilon^{ijmn}\,K_{mn}{}^{kl} = \epsilon^{kjmn}\,K_{mn}{}^{il} +
\epsilon^{ikmn}\,K_{mn}{}^{jl}+ \epsilon^{ijkn}\,K_{mn}{}^{ml} +
\epsilon^{ijmk}\,K_{mn}{}^{nl},\label{Kid2}\\
\epsilon^{ijmn}\,K_{mn}{}^{lk} = \epsilon^{ljmn}\,K_{mn}{}^{ik} +
\epsilon^{ilmn}\,K_{mn}{}^{jk}+ \epsilon^{ijln}\,K_{mn}{}^{mk} +
\epsilon^{ijml}\,K_{mn}{}^{nk}.\label{Kid3}
\end{eqnarray}
Suppose that the tensor $K$ is {\it traceless}: $K_{mn}{}^{mk} = 0$.
Then using (\ref{Kid2})-(\ref{Kid3}) in (\ref{Kid1}), we find:
\begin{equation}
\epsilon^{ijmn}\,K_{mn}{}^{[kl]} = \epsilon^{jmn[k}\,K_{mn}{}^{l]i} -
\epsilon^{imn[k}\,K_{mn}{}^{l]j}.\label{Kid4}
\end{equation}
Now we once again apply the identity (\ref{Zident}):
\begin{equation}
\epsilon^{jmnk}\,K_{mn}{}^{li} = \epsilon^{jmnl}\,K_{mn}{}^{ki}
+ \epsilon^{jmlk}\,K_{mn}{}^{ni} + \epsilon^{jlnk}\,K_{mn}{}^{mi}
+ \epsilon^{lmnk}\,K_{mn}{}^{ji}.\label{Kid5}
\end{equation}
Taking into account the tracelessness, $K_{mn}{}^{mk} = 0$, we can rearrange
the terms (move the first term from the r.h.s.\ to the l.h.s.) and find
\begin{equation}\label{Kid6}
\epsilon^{jmn[k}\,K_{mn}{}^{l]i}={\frac 1 2}\,\epsilon^{klmn}\,K_{mn}{}^{ij}.
\end{equation}
Finally, using (\ref{Kid6}) in (\ref{Kid4}), we prove the {\bf
  K-identity}: {\it Every (2,2) tensor $K_{ij}{}^{kl}$ that is
  traceless $K_{ki}{}^{kj} = 0$ has the property}
\begin{equation}\label{Kident}
\epsilon^{ijmn}\,K_{mn}{}^{[kl]} = \epsilon^{klmn}\,K_{mn}{}^{[ij]}.
\end{equation}

[Incidentally, this property applies in particular to the Weyl
curvature tensor $C_{ij}{}^{kl}$. In this case, we recover from
(\ref{Kident}) the well known anti-self double-duality of the Weyl
tensor: $C_{ij}{}^{kl} = {\frac 1 4}\,\epsilon^{klmn}
\,\hat{\epsilon}_{ijpq}\,C_{mn}{}^{pq}$].

\end{document}